\input harvmac
\input epsf

\def\rhob{{\rho\kern-0.465em \rho}}

\def\o{\over}

\def\ontopss#1#2#3#4{\raise#4ex \hbox{#1}\mkern-#3mu {#2}}

\setbox\strutbox=\hbox{\vrule height12pt depth5pt width0pt}

\def\strut{\relax\ifmmode\copy\strutbox\else\unhcopy\strutbox\fi}


\nref\rbnw{V.V. Bazhanov, B. Nienhuis, and S.O. Warnaar, 
Phys. Lett. B 322 (1994) 198.}
\nref\rwnsa{S. O. Warnaar, B. Nienhuis and K.A. Seaton,
Phys. Rev. Lett. 69 (1992) 710.}
\nref\rwnsb{S. O. Warnaar, B. Nienhuis and K.A. Seaton,
Int. J. Mod. Phys. B 7 (1993) 3727.}
\nref\rbr{V.V. Bazhanov and N.~Reshetikhin, Prog. Theo. Phys. Suppl. 102
(1990) 301.}
\nref\rkm{T.R. Klassen and E. Melzer, Nucl. Phys. B338 (1990) 485.}
\nref\ralzam{Al.B. Zamolodchikov, Phys. Lett. B253 (1991) 391.}
\nref\rzama{A.B. Zamolodchikov, Adv. Stud. in Pure Math. 19 (1989) 641.}
\nref\rzamb{A.B. Zamolodchikov, Int. J. Mod. Phys. A4 (1989) 4235.}
\nref\rkkmm{R. Kedem, T.R. Klassen, B.M. McCoy and E. Melzer,
Phys. Letts. B 304 (1993) 263.}
\nref\rkmm{R. Kedem, B.M.McCoy and E. Melzer, in {\it Recent Progress
in Statistical Mechanics and Quantum Field Theory} ed. P. Bouwknegt et
al. (World Scientific, Singapore, 1995) 195.}
\nref\rwp{S.O. Warnaar and P.A. Pearce, J. Phys. A27 (1994) L891.}
\nref\rgna{U. Grimm and B. Nienhuis,{\it Scaling properties of the
Ising model in a field}, Proceedings of the seventh Nankai Workshop
(Tianjin 1995), ed. M.-L. Ge and F.Y. Wu.}
\nref\rgnb{U. Grimm and B. Nienhuis, {\it Scaling limit of the Ising
model in a field}, Phys. Rev. E (to be published).}
\nref\radm{G. Albertini, S. Dasmahapatra and B.M. McCoy,
Phys. Letts. A 170 (1992) 397.}
\nref\rbcds{H.W. Braden, E. Corrigan, P.E. Dorey and R. Sasaki,
Nucl. Phys. B338 (1990) 689.}
\nref\rmckay{J. McKay, Proc. Symp. Pure Math. 37 (1980) 183.}
\nref\rkost{B. Kostant, Proc. Natl. Acad. Sci. USA 81 (1984) 5275.}
\nref\rdorey{P.E. Dorey, in {\it Integrable Quantum Field Theories} ed.
L. Bonuro et al., Plenum (1993), (NATO Advanced Study Institute,
Series B; Physics V. 310), 83.}
\nref\rpas{V. Pasquier, Nucl. Phys. B285 [FS 19](1987) 162.}

\Title{\vbox{\baselineskip12pt
  \hbox{ITPSB 96-63}}}
  {\vbox{\centerline{Single Particle Excitations in the Lattice 
  $E_8$ Ising Model}}}


  \centerline{ Barry~M.~McCoy\foot{mccoy@max.physics.sunysb.edu} and
               William~P.~Orrick\foot{worrick@insti.physics.sunysb.edu}}

  \bigskip\centerline{\it Institute for Theoretical Physics}
  \centerline{\it State University of New York}
  \centerline{\it Stony Brook,  NY 11794-3840}
  \bigskip
  \Date{\hfill 11/96}

  \eject

\centerline{\bf Abstract}
 We present analytic expressions for the single particle 
excitation energies of the 8 quasi-particles in the lattice $E_8$ 
Ising model and demonstrate that all excitations have an extended
Brillouin zone which, depending on the excitation, ranges from
$0<P < 4\pi$ to $0< P< 12 \pi.$ These are compared with
exact diagonalizations for systems through size 10 and with the $E_8$
fermionic representations of the characters of the critical system 
 in order to study the counting statistics. 

\newsec{Introduction}
 Several years ago Bazhanov, Nienhuis and Warnaar~\rbnw~demonstrated
that the the solution of the  dilute $A_3$ model of Warnaar, Nienhuis and
Seaton~\rwnsa-\rwnsb~may be reduced to the Bethe's Ansatz equation based
on the group $E_8$ given in~\rbr-\ralzam. This thus provides a lattice
realization of the $E_8$
continuum field theory found by Zamolodchikov~\rzama-\rzamb~ in 1989 
to be in the
same universality class as the Ising model in a magnetic field at $T=T_c.$

The reduction in~\rbnw~to the
$E_8$ equations is based on a conjecture for the allowed string type
solutions of the Bethe's Ansatz equations for the dilute $A_3$ model and this
conjecture was motivated by a numerical study of the system at
criticality in the sector $l=1$ which contains the ground
state. This study gives a set of counting rules for the spectrum which
exactly reproduces the $E_8$ fermionic representation of the character
$\chi_{1,1}^{(3,4)}(q)$ of the Ising model which was conjectured in
~\rkkmm-\rkmm~and proven in its polynomial generalization in ~\rwp.

However, there is as yet no satisfactory conjecture for the string
content at criticality of all
states in the sectors $l=2$ and $l=3$ (which correspond to the
characters $\chi_{1,2}^{(3,4)}(q)$ and $\chi_{1,3}^{(3,4)}(q)$) 
and a recent study~\rgna-\rgnb~of numerical
solutions to the non-critical Bethe's equations at zero momentum
indicates that the string identification may at times change as the
system moves away from criticality. We make a two fold study of these
questions here by numerically computing the single particle
excitations for finite chains of sizes up through 10 and by comparing
these results with the single particle dispersion relations which are
obtained by extending the computations of~\rbnw. We find that all
eight quasi-particles exhibit the phenomenon of an extended Brillouin
zone such as was first explicitly seen in the anti-ferromagnetic 
three state Potts spin chain~\radm~and
that there are no genuine one particle states in the two sectors
$l=2$ and $3.$

We present the dispersion relations in Sec.~2, the
numerical study in Sec.~3 and discuss our findings in Sec.~4. We
follow the notation of ref.~\rbnw~and refer the reader to that paper
for the explicit Boltzmann weights which define the model. We will here
be concerned only with the Hamiltonian version of the model.

\newsec{Single Particle Dispersion Relations and Characters}

In~\rbnw~and~\rbr~the single particle excitations were studied in the limit
$L\rightarrow \infty$ and in particular
the Fourier transform of the single particle excitations in
sector $l=1$ is
computed. We have inverted the transform to find the results given
below. The computations involve  some tedious algebra and moreover for
the non-critical case the required integrals do not seem to be in the
literature. The details will be given elsewhere.

For the critical case we have as a function of rapidity $u$
\eqn\disp{\eqalign{
P_{j}(u)&=\sum_a\left[{\pi}+2 \arctan\left({\sinh u\over
\sin(a\pi/30)}\right)\right]\cr
e_{j}(u)&=A{d P_{j}(u)\over du}=A \cosh u\sum_a{\sin (a\pi/30)\over \cosh^2 u
-\cos^2 (a\pi/30)}\cr}}
where $A=32/15$ , $-\infty<u <\infty,$ the number thirty has
the significance of being the Coxeter number of $E_8$ and  
$a$ takes on the following  values  
as a function of $j$ (we use the labeling of the Cartan matrix $E_8$
shown in Fig. 1 and indicate the identification of particles 
of ~\rbnw~in parentheses)
\eqn\ifora{\eqalign{j~~~~~~~~&a~~~~~~~~~~~~~~~~~~~~~~\Delta P\cr
                 1~~(1)~~~~~&1,~11~~~~~~~~~~~~~~~~~~4\pi\cr
                 2~~(7)~~~~~&7,13~~~~~~~~~~~~~~~~~~~4\pi\cr
                 3~~(2)~~~~~&2,10,12~~~~~~~~~~~~~~~6\pi\cr
                 4~~(8)~~~~~&6,10,14~~~~~~~~~~~~~~~6\pi.\cr
                 5~~(3)~~~~~&3,9,11,13~~~~~~~~~~~~8\pi\cr
                 6~~(6)~~~~~&6,8,12,14~~~~~~~~~~~~8\pi\cr
                 7~~(4)~~~~~&4,8,10,12,14~~~~~~~10\pi\cr
                 8~~(5)~~~~~&5,7,9,11,13,15~~~~12\pi\cr}}
The numbers $a$ in this table agree with the corresponding numbers for the
scattering of a particle of type 1 with a particle of one of the 8 types
of the $E_8$ field theory as given in~\rbcds~and have group theoretical
and geometrical interpretations~\rmckay-\rdorey.
They also follow from (4.19) of~\rbnw. The numbers $\Delta P$ are
$2\pi$ times the elements in the first column of the inverse Cartan matrix.

For the non-critical case in which the nome $q_B$ of the Boltzmann weights 
of~\rbnw~(which is essentially
the magnetic field) is not zero we have the following generalization 
which reduces to~\disp~when  $q\rightarrow 0$ (and the modulus
$k\rightarrow 0$)
\eqn\masdisp{\eqalign{
P_{j}(u,q)&=\sum_{a}\left[\pi+2 \arctan \left({i\ {\rm sn}
\ iu\over {\rm sn}(a  K/15)}\right)\right]\cr
e_{j}(u,q)&=A(q){d P_{j}(u)\over du}=A(q)\ {\rm dn}iu \ {\rm cn}\ iu
\sum_{a}{{\rm sn}(a K/15)\over 
{\rm cn}^2 iu -{\rm cn}^2( a K/15)}\cr}}
where $q=q_B^{16/15}$, $A(q)=2A K(q)/\pi$,  $K(q)$ is the complete
elliptic integral of the first kind and $-K'<u<K'.$
This expression is in fact a universal form for all the Bethe's Ansatz
models based on a simply laced Lie algebra, the only difference 
being that the values of $a,$ here given by~\ifora,~vary
from model to model.

The minimum in the energy occurs $u=\pm K'$ where $P=0$ (or the value
$\Delta P$ given in ~\ifora). Thus we find
\eqn\emin{e_{j}(P=0,q)=kA(q)\sum_{a}{\rm sn}(aK/15)}
and in particular
\eqn\zerolim{\lim_{k\rightarrow 0}e_{j}(P=0,q)/kA(q)=\sum_a\sin(a\pi/30).}
It may be verified that these values coincide with the components of
the Perron-Frobenius eigenvector of the $E_8$ Cartan matrix (as given,
for example, in~\rpas.)

When $q\rightarrow 1~(k\rightarrow 1)$ we find that
\eqn\elim{e_{j}(u,q)\rightarrow A(q)\sum_a 1}
where the normalizing constant is diverging. Thus using ~\ifora~
 we see that the
single particle states for particles 1 and 2 become degenerate as
$q\rightarrow 1$ (and similarly for particles 3 and 4). For the
remaining particles  degeneracies with multi-particle states also occur.

In principle $u$ can be eliminated between the two expressions
in~\masdisp~to produce a polynomial relation between $e_j$ and $P_j.$
However here
we have done the elimination  numerically and present
the results
in Fig.~2 for $q_B=0$ and in Fig. 3 for $q_B=0.2$. In Fig.~3~we
see that
the labeling we have used corresponds to the ordering of eigenvalues
at $P=0$ which is the same for all $q>0.$  
It is to be explicitly remarked that the dispersion relations do not
have the restriction $0 <P<2\pi$ but in all cases have the larger
momentum range $0<P<\Delta P$ where $\Delta P$ is given in \ifora. 
This is the phenomenon of the extended Brillouin zone
scheme. This is a very general phenomenon in integrable models first explicitly
seen in~\radm~for the three state anti-ferromagnetic Potts spin chain.

We also wish  to make contact with the characters of conformal field
theory and finite size computations.
The fermionic representation of characters of the Ising model
in the $E_8$ basis is given~\rkkmm-\rkmm~ in terms of the fermionic form
\eqn\ferfor{\sum_{n_1,\ldots ,n_8=0}^{\infty} ~
    {q^{{\bf n} C_{E_8}^{-1} {\bf n}-{\bf A}\cdot{\bf n} }
 \o (q)_{n_1} \ldots (q)_{n_8} },}
where $(q)_n=\prod_{j=1}^n(1-q^j),$
 $C_{E_8}$ is the Cartan matrix of the Lie algebra $E_8$
given by the incidence matrix of Fig.~1 (where we use the labeling of
~\rkkmm) and we trust that $q$ will not be confused with the nome of
the elliptic functions.
Explicitly
\eqn\cinv{C_{E_8}^{-1}=\pmatrix{2& 2& 3& 3& 4& 4& 5& 6 \cr
                                          2& 4& 4& 5& 6& 7& 8& 10 \cr
                                          3& 4& 6& 6& 8& 8& 10& 12 \cr
                                          3& 5& 6& 8& 9& 10& 12& 15 \cr
                                          4& 6& 8& 9& 12& 12& 15& 18 \cr
                                          4& 7& 8& 10& 12& 14& 16& 20 \cr
                                          5& 8& 10& 12& 15& 16& 20& 24 \cr
                                          6& 10& 12& 15& 18& 20& 24& 30 \cr}}
For the characters $\chi_{1,s}^{(3,4)}(q)$ (normalized to
$\chi_{1,s}^{(3,4)}(0)=1$) it was conjectured
in~\rkkmm~and proven in~\rwp~that
\eqn\chioo{\chi_{1,1}^{(3,4)}(q)=\sum_{n_1,\ldots ,n_8=0}^{\infty} ~
    {q^{{\bf n} C_{E_8}^{-1} {\bf n}}
 \o (q)_{n_1} \ldots (q)_{n_8} }}
and in~\rkmm~it was conjectured on the basis of computer studies that
\eqn\cot{\eqalign{
\chi_{1,1}^{(3,4)}(q)+\chi_{1,2}^{(3,4)}(q)&
=\sum_{n_1,\ldots ,n_8=0}^{\infty} ~
    {q^{{\bf n} C_{E_8}^{-1}  {\bf n}-{\bf A}^{(1)}\cdot{\bf n} }
 \o (q)_{n_1} \ldots (q)_{n_8} }~~{\rm with}~~A_j^{(1)}
=(C_{E_8}^{-1})_{1,j}\cr
\chi_{1,1}^{(3,4)}(q)+\chi_{1,2}^{(3,4)}(q)+
\chi_{1,3}^{(3,4)}(q)&=\sum_{n_1,\ldots ,n_8=0}^{\infty} ~
    {q^{{\bf n} C_{E_8}^{-1} {\bf n}-{\bf A}^{(2)}\cdot{\bf n} }
 \o (q)_{n_1} \ldots (q)_{n_8} }~~
{\rm with}~~A^{(2)}_j=(C_{E_8}^{-1})_{2,j}.\cr}}

We also note that 
the character~\chioo~is derived  from the lattice quasiparticle
spectrum
\eqn\equasi{E-E_{GS}=\sum_{j=1}^8~\sum_{i=1}^{n_j}
e_{j}(P_j^{i})}
\eqn\pquasi{P=\sum_{j=1}^8~\sum_{i=1}^{n_{j}}
  P_j^{i}~~({\rm mod}~2\pi),}
where $e_{j}(P)$ are single particle dispersion relations, we have the
fermionic restriction
\eqn\resf{P_j^{i}\neq P_j^{k}~~~{\rm for }~~i\neq k~~{\rm and~all}~ j}
 and the momenta are chosen from the set
\eqn\mom{P_{j}^{i} \in \Bigl\{ P_{j}^{\rm min}({\bf n}), 
  ~P_{j}^{\rm min}({\bf n})+{2\pi \o L},
  ~P_{j}^{\rm min}({\bf n})+{4\pi \o L},
  ~\ldots,
  ~P_{j}^{\rm max}({\bf n}) \Bigr\}~,}
with
\eqn\Pmin{P_{j}^{\rm min}({\bf n})
  ~ =~{2\pi \over L}~ \Bigl[ ({\bf n}C_{E_8}^{-1})_j+{1\over
2}(1-n_j) \Bigr]}
and
\eqn\Pmax{ P_{j}^{\rm max}({\bf n}) ~=~ -P_{j}^{\rm min}({\bf n})+
                  {2\pi (C_{E_8}^{-1})_{1,j}~.}}

\newsec{Finite Size Study}

We have numerically determined the Hamiltonian eigenvalues for
chains up through size 10.  We have determined the sector $l$ from the
property that at $q=0$ eigenvectors in the sector $l=2 (l=1,3)$ are
antisymmetric (symmetric) under the interchange $1 \leftrightarrow 3$
of the states in the basis of~\rwnsa,~and by a similar symmetry property
under the interchange $2 \leftrightarrow (1+3)/\sqrt{2}$ that 
distinguishes $l=1$ from $l=3$.
We present these data in Figs.~4--7 where we also compare with the
$L\rightarrow \infty$ formula~\masdisp.

In Fig.~4 we plot at $q=0$ the eigenvalues which behave as $2A(q)$ as
$q\rightarrow 1.$ These include all the single particle states for
particles 1 and 2 in the sector $l=1$ which are allowed by
~\mom. These states are indicated with triangles. The remaining states
are to be found in the sectors $l=2$ and $l=3.$ We note that while the
number of states in $l=1$ grows proportionally with $L$ that there are
only 4 states with $l=2$ and 2 states with $l=3.$  The two $l=2$ states
of particle 1 at $P=0,2\pi/10$ are accounted  
for by subtracting~\chioo~from the first equation of~\cot~to write
\eqn\first{\chi_{1,2}^{(3,4)}(q)
=\sum_{n_1,\ldots ,n_8=0}^{\infty} ~
    {q^{{\bf n} C_{E_8}^{-1}  {\bf n}-{\bf A}^{(1)}\cdot{\bf n}} 
    -q^{{\bf n} C_{E_8}^{-1}  {\bf n}}
\o (q)_{n_1} \ldots (q)_{n_8} }}
and by considering the $n_1=1,n_2,\ldots,n_8=0$ term
of the sum to find $(1-q^2)/(1-q)=1+q.$
Similarly, the two $l=2$ states of particle 2 at $P=4\pi/10,6\pi/10$ 
are obtained from ~\first  by setting  $n_1=0,n_2=1,n_3,\ldots,n_8=0$ 
to obtain $(q^2-q^4)/(1-q)=q^2+q^3.$
and  the two $l=3$ states of particle 2 at $P=0,2\pi/10$ are
obtained by first using ~\cot to write
\eqn\four {\chi_{1,3}^{(3,4)}(q)
=\sum_{n_1,\ldots ,n_8=0}^{\infty} ~
    {q^{{\bf n} C_{E_8}^{-1}  {\bf n}-{\bf A}^{(2)}\cdot{\bf n}} 
     -q^{{\bf n} C_{E_8}^{-1}  {\bf n}-{\bf A}^{(1)}\cdot{\bf n}}
\o (q)_{n_1} \ldots (q)_{n_8} }}
and then setting  $n_1=0,n_2=1,n_3,\ldots,n_8=0$ 
to obtain $(1-q^2)/(1-q)=1+q.$

In Fig.~5 we do a similar study for the states which behave as $4A(q)$
as $q\rightarrow 1.$ The states which are the single particle
excitations for particles 3 and 4 in $l=1$ are again marked with
triangles and of the remaining states there are 9 in  $l=2$ and 5 in $l=3.$
As before, the character representations~\first~and~\four~account for the
$l=3$ state of particle 3 at $P=4\pi/10$, the $l=2$ states of particle 3 at
$P=6\pi/10,8\pi/10,\pi$, the $l=3$ states of particle 4 at $P=6\pi/10,8\pi/10$
and the $l=2$ states of particle 4 at $P=\pi,12\pi/10,14\pi/10$.

We see from Figs.~4 and~5~that the states with $l=2$ and $l=3$ exactly fill
up the states missing in
$l=1.$ We also see that the $L\rightarrow \infty$ excitation curves
give an excellent fit to the finite size energies for $l=1$ for all
momenta which are allowed by~\mom. Such agreement is not particularly
seen for the remaining states in $l=2,3.$ 

In Figs.~6 and~7 we study this same selection of states for $q>0$ and
compare with the massive dispersion relation~\masdisp. Here we see that
if $q$ is sufficiently large then all states, regardless of their
sector, lie on the same massive dispersion curve.

\newsec{Conclusions}

We may now discuss the findings of~\rgna-\rgnb~and the relation with~\rbnw.
The study in~\rgna-\rgnb~is restricted to $P=0$ and the major
conclusions are 1) that the string content of the $P=0$ states
qualitatively changes with $q$ and 2) that the masses of particles $1-5$ as
determined from $P=0$ agree with the masses of the $E_8$ field
theory of~\rzama-\rzamb. These conclusions are in agreement with our
data. On the other hand it is perhaps not exactly fair to say that the
single particle states 1 and 3 lie in sector $l=2$ and particles 2
and 4 lie in sector $l=3$ because the {\it only} sector at $q=0$ which
contains single particle states is $l=1.$ When $q$ increases there is a
motion of the finite number of states in $l=2,3$ to smoothly join the
macroscopic number of states in $l=1.$ This is the numerical explanation of why
the mass computations done in~\rbnw~in the sector $l=1$ are correct
even though $P=0$ is never in the sector $l=1.$ This motion is seen
numerically  in~\rgna-\rgnb~where the crossover happens on a scale of
$q\sim L^{-15/8}.$ It is clearly most
desirable to prove this analytically.

We note that the absence of genuine single particle states in
$l=2,3$ means that these sectors have a somewhat different nature
than $l=1$ and it is presumably for this reason that a representation
of the spectrum in the form~\pquasi--\Pmax~has not yet been
found. Such a representation is needed to derive and interpret the
character formulas~\cot.

We also note that the recognition that the excitations all have an
extended Brillouin zone scheme renders  obsolete the suggestion in ~\rgnb~that 
the model contains
massive particles at  criticality.

Finally we remark that the question must be raised as to what the
introduction of non-integrability (no matter how small) will have on the
single particle excitations. In particular the pieces of the spectrum
in the extended Brillouin zone which lie far up in the spectrum and cross a
very large number of levels when $q\rightarrow 0$ should be expected
to decay in some fashion when non-integrability is introduced. This
should be relevant to the relation of the  $E_8$ integrable model to
the ordinary Ising model in a magnetic field at $T=T_c$ and remains to
be explored.

\bigskip
{\bf Acknowledgments}

We are indebted to B. Nienhuis, U. Grimm and S.O.Warnaar for extensive
discussions and for providing us with unpublished material and to G. Mussardo
and P.E. Dorey for useful conversations.

This work is supported in part by the NSF under DMR9404747.

\vfill
\eject
 \centerline{\epsfxsize=3in\epsfbox{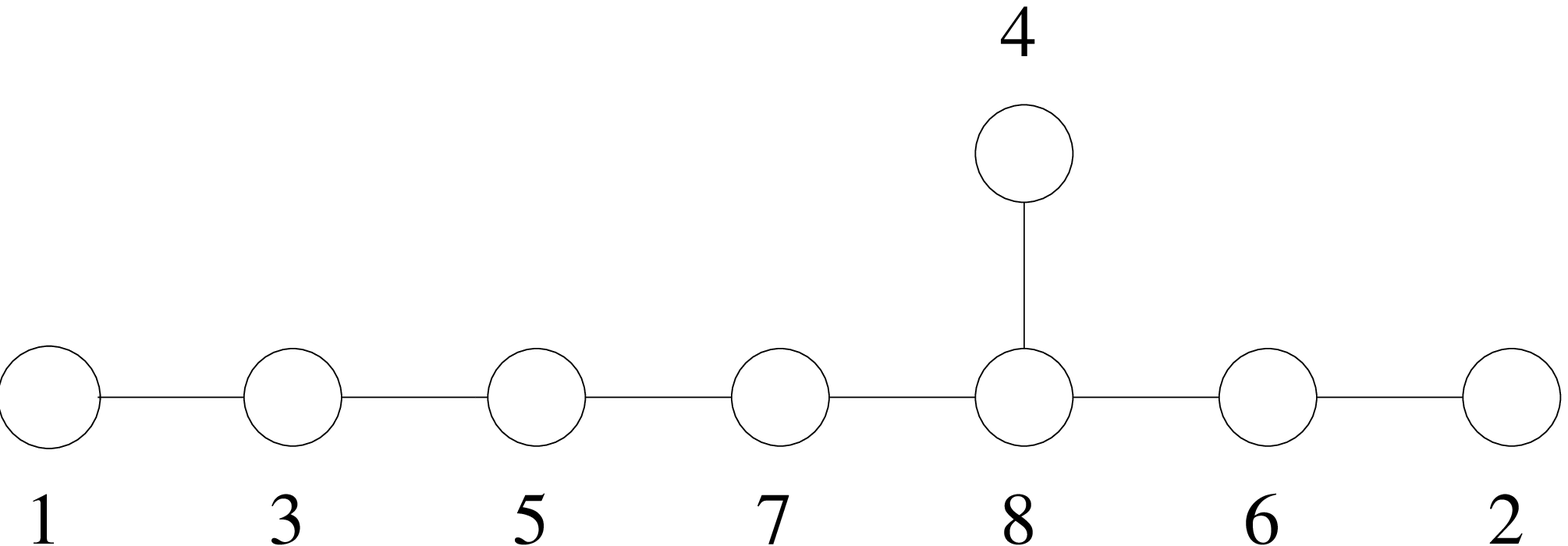}}
Fig. 1. The Dynkin diagram for the Lie group $E_8$ with the labeling
of nodes used in this paper.
\vfill                 
\eject
 \centerline{\epsfxsize=6in\epsfbox{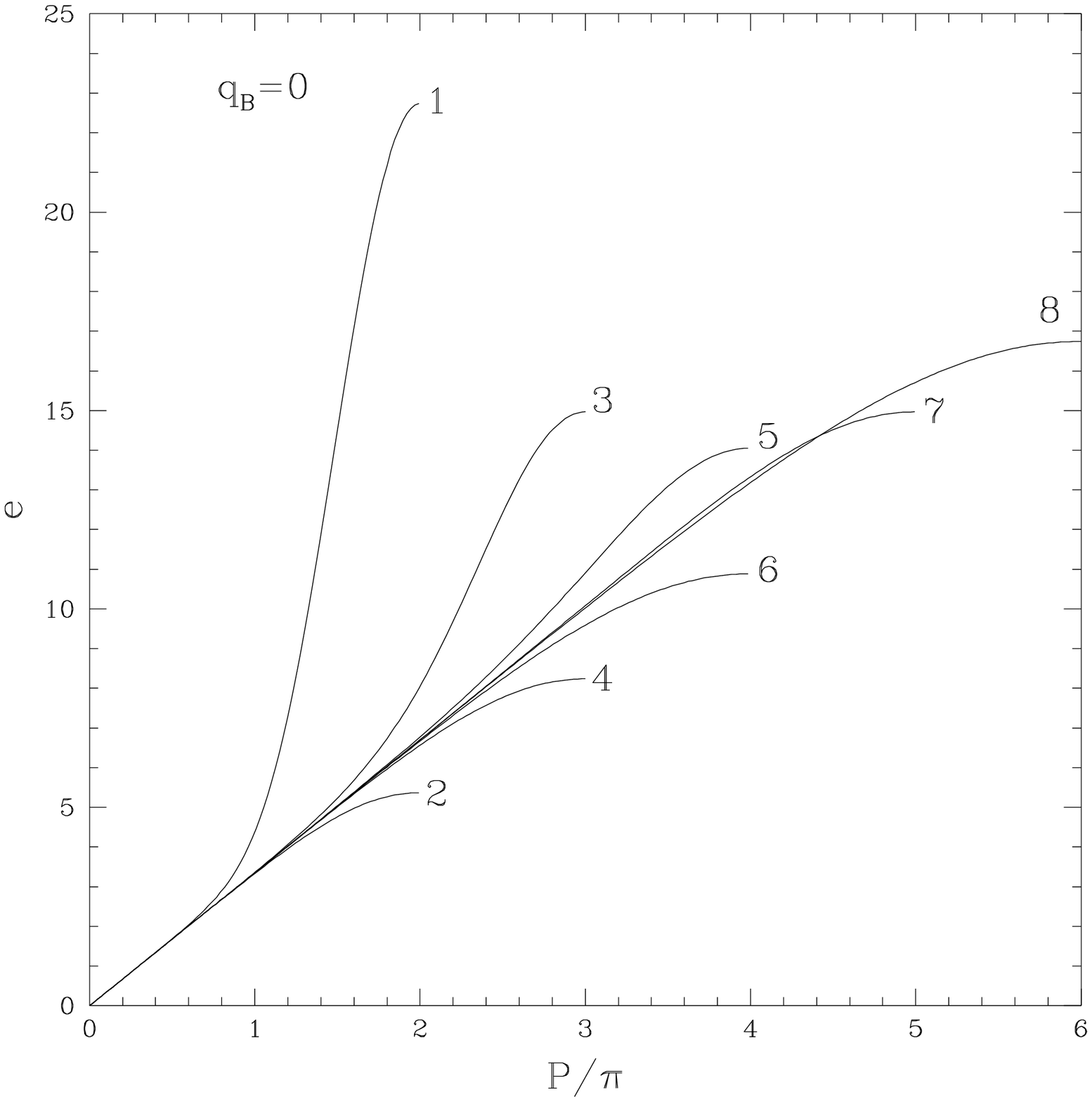}}
Fig. 2. The eight single particle dispersion relations at $q_B=0.$
\vfill                 
\eject
 \centerline{\epsfxsize=6in\epsfbox{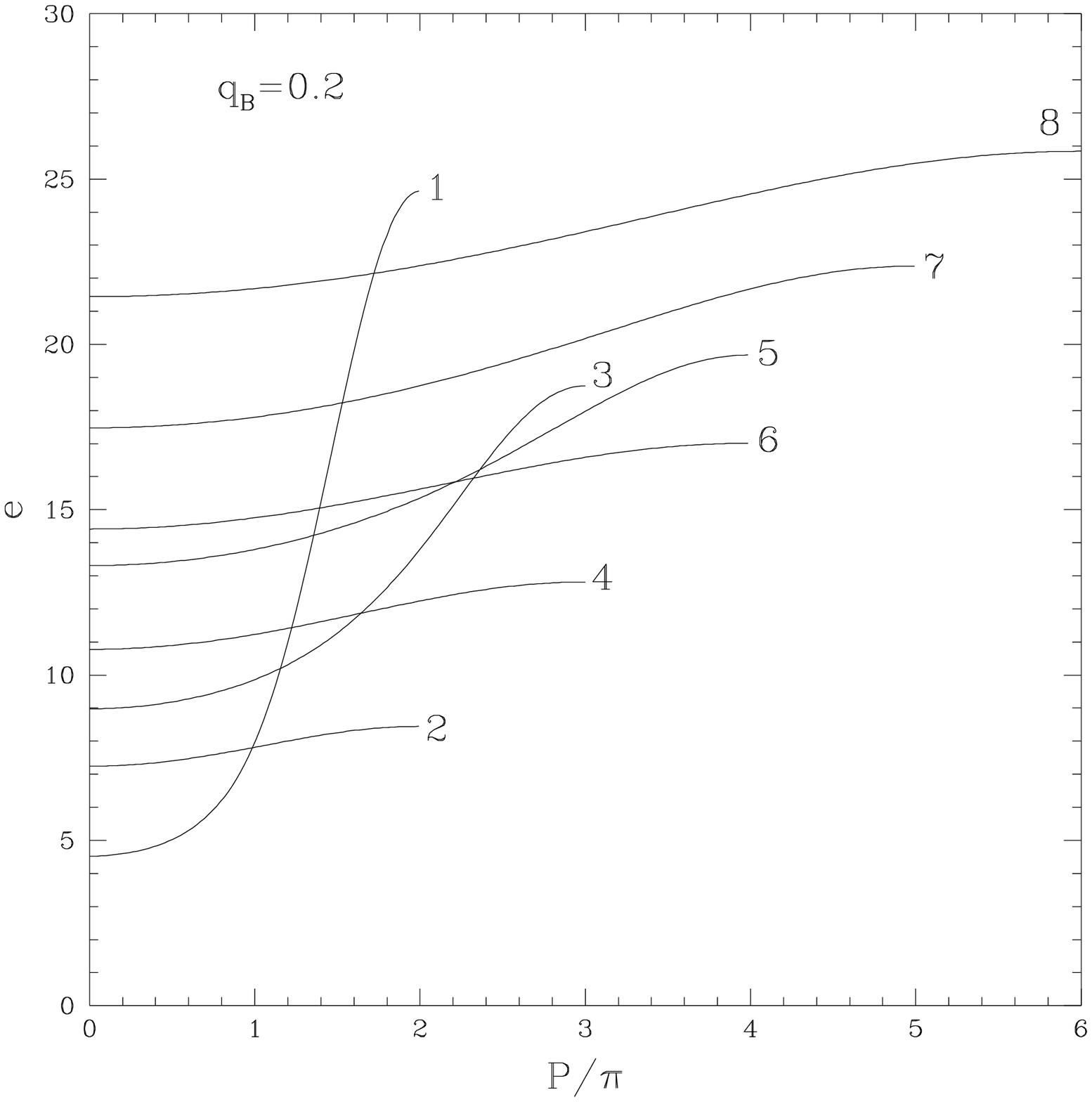}}
Fig. 3. The eight single particle dispersion relations at $q_B=0.2.$ 
\vfill                 
\eject
 \centerline{\epsfxsize=6in\epsfbox{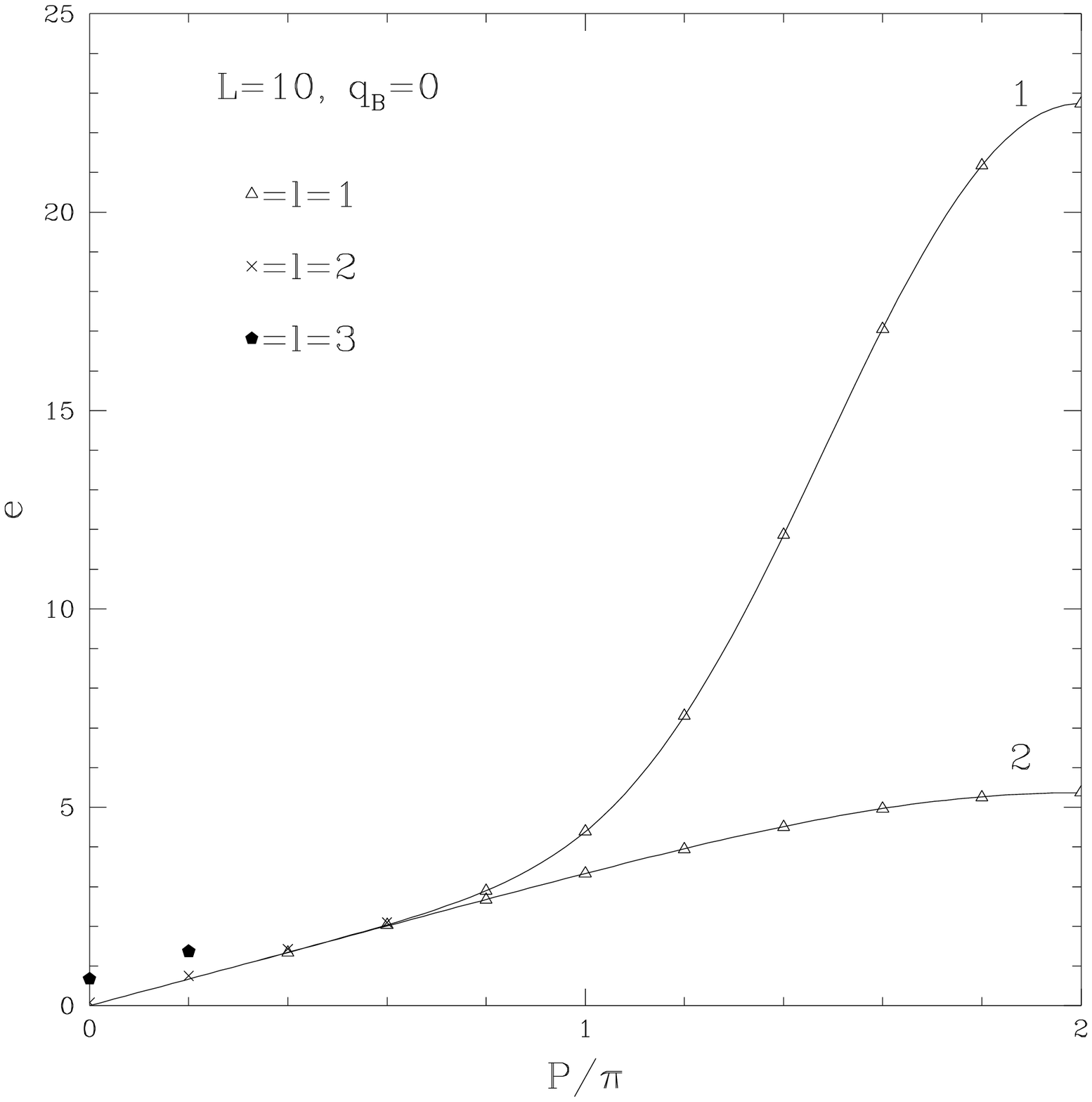}}

Fig. 4. The plot for $q_B=0$ and $L=10$ of the states which for 
$q\rightarrow 1$ behave as $2A(q).$ The states in $l=1$ are marked by
triangles, in $l=2$ by crosses and in $l=3$ by pentagons. The smooth
curves are the theoretical curves~\disp~for particles 1 and 2 for $l=1$ 
and $L\rightarrow \infty.$ Note that particle 1 always lies above
particle 2.

\vfill                 
\eject
 \centerline{\epsfxsize=6in\epsfbox{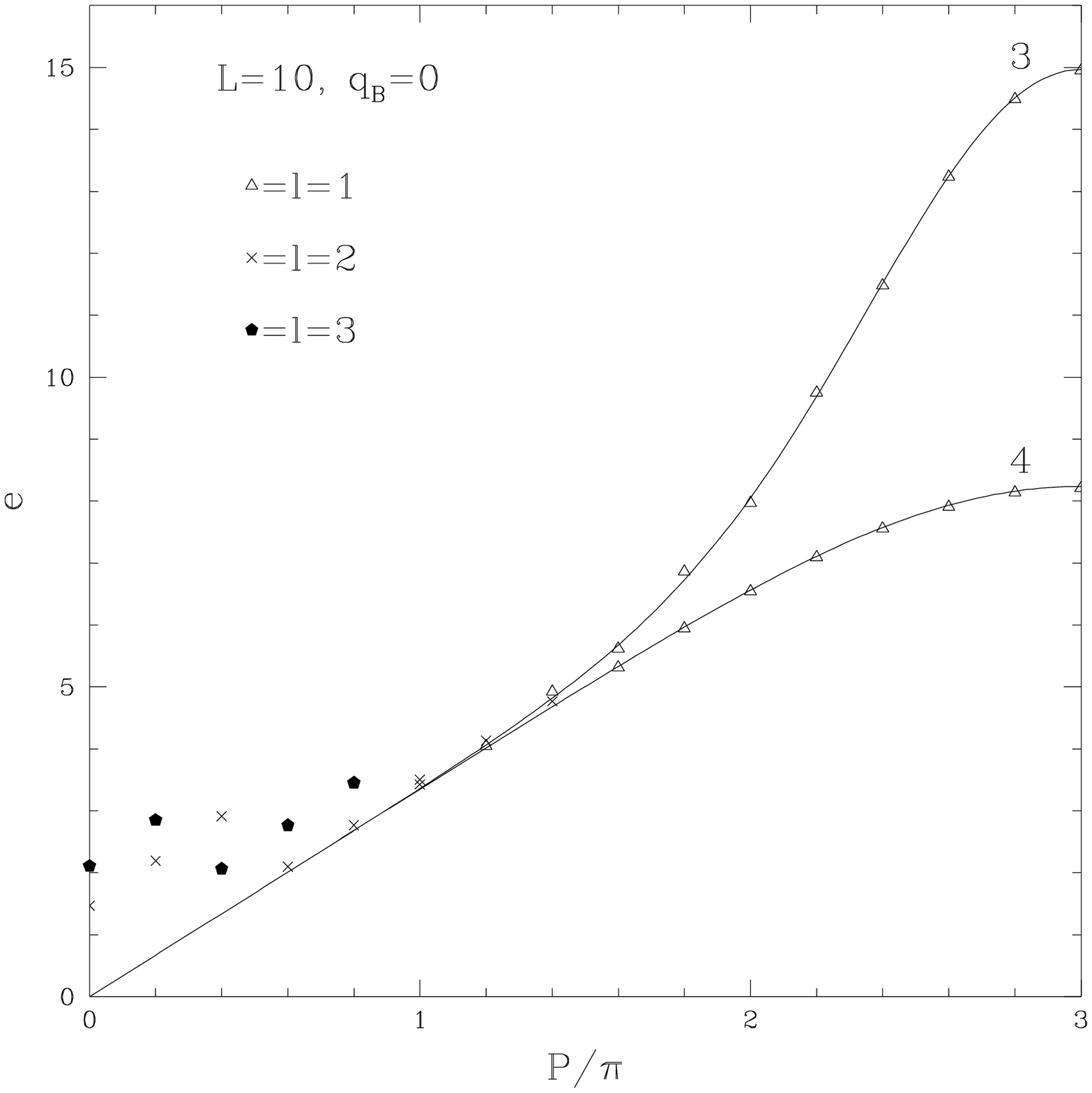}}

Fig. 5.   The plot for $q_B=0$ and $L=10$ of the states which for
$q\rightarrow 1$ behave as $4A(q).$ The states in $l=1$ are marked by
triangles, in $l=2$ by crosses and in $l=3$ by pentagons.  The smooth
curves are the theoretical curves~\disp~for particles 3 and 4 for $l=1$
and $L\rightarrow \infty.$
Note that particle 3 always lies above particle 4. 

\vfill                 
\eject
 \centerline{\epsfxsize=6in\epsfbox{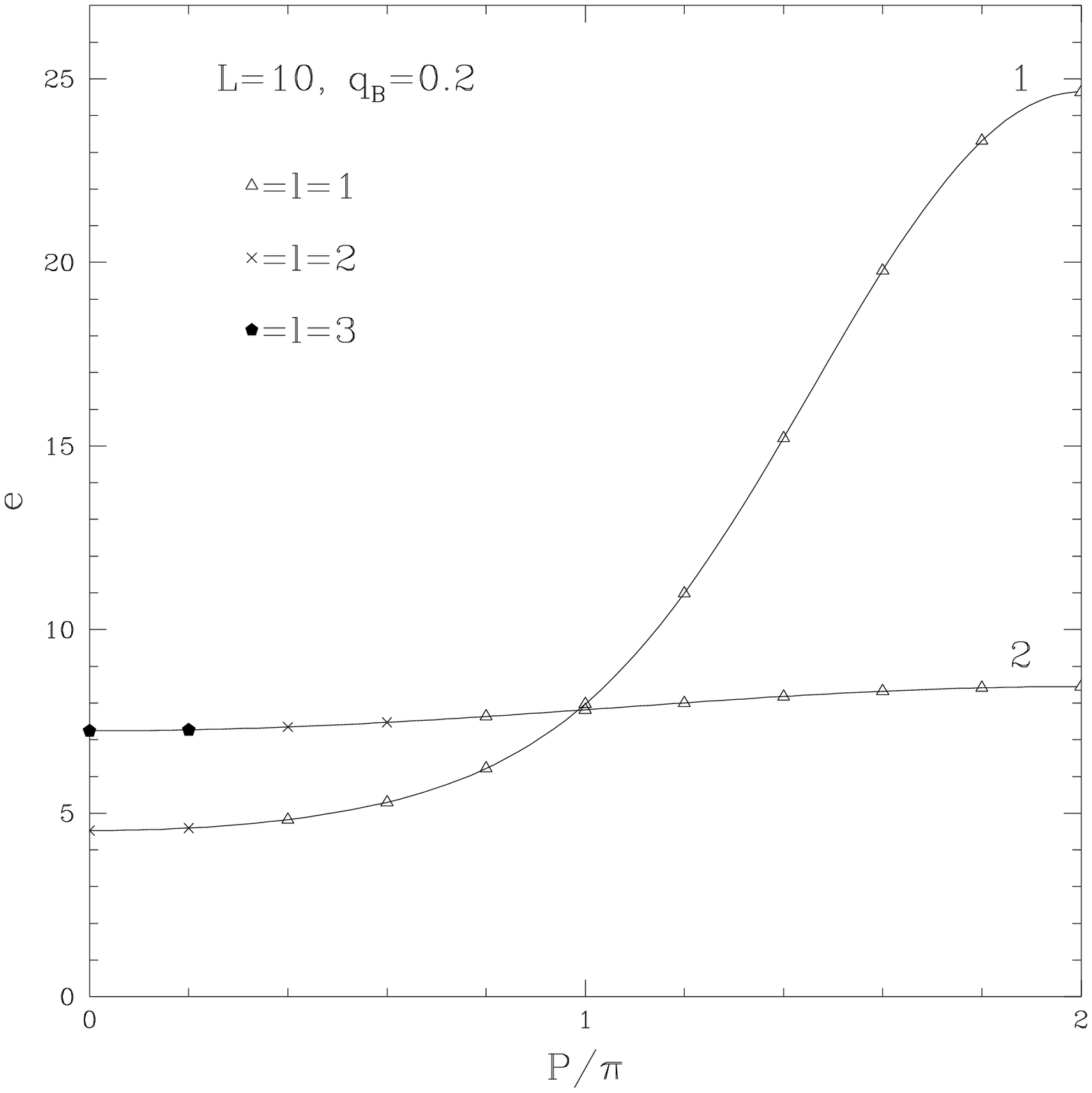}}

Fig. 6.     The plot for $q_B=0.2$ and $L=10$ of the states which for
$q\rightarrow 1$ behave as $2A(q).$ The states in $l=1$ are marked by
triangles, in $l=2$ by crosses and in $l=3$ by pentagons. The smooth
curves are the theoretical curves~\masdisp~for particles 1 and 2 for $l=1$ 
and $L\rightarrow \infty.$ Note that these two curves cross near $P=\pi.$ 

\vfill                 
\eject
 \centerline{\epsfxsize=6in\epsfbox{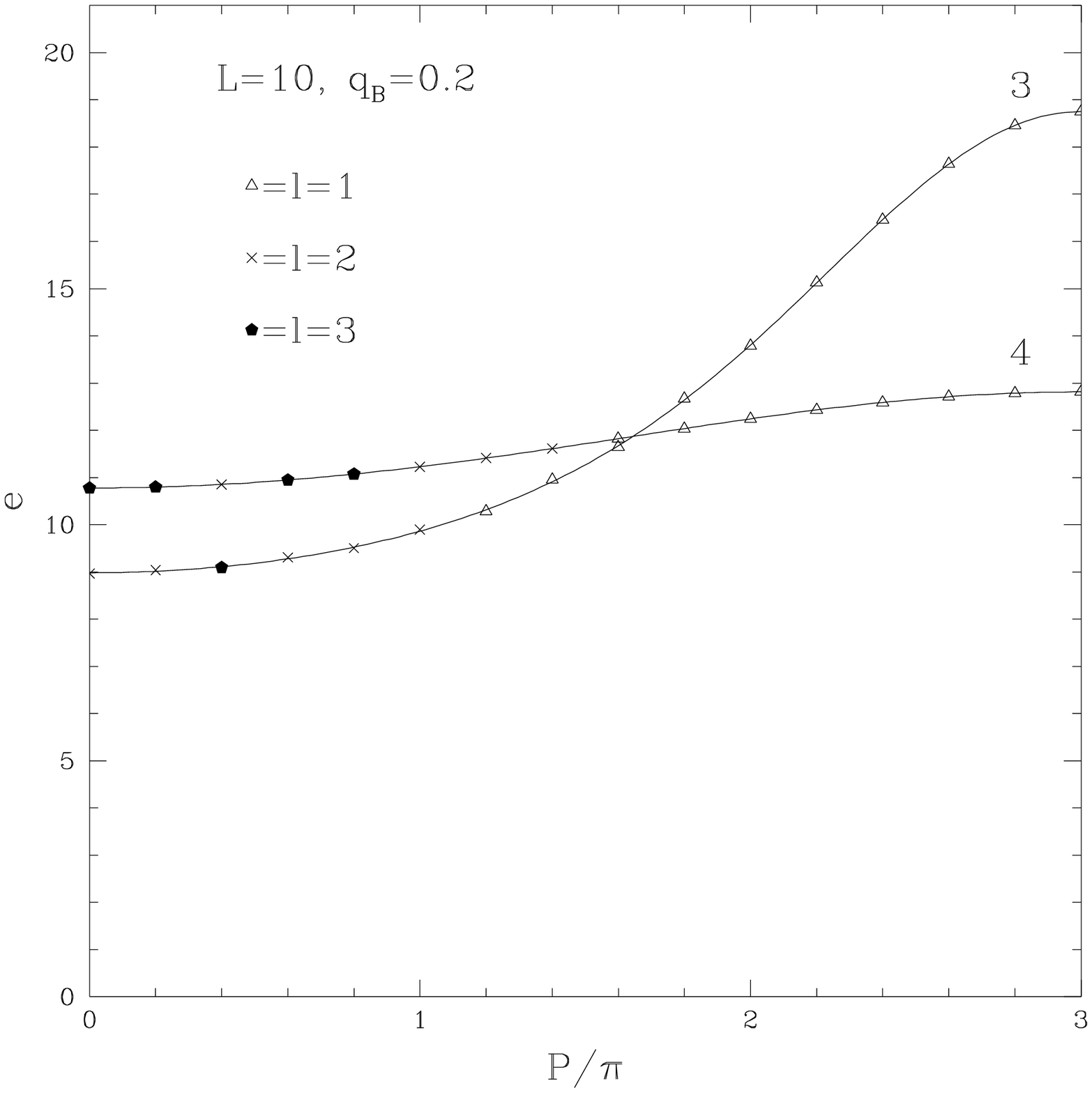}}

Fig. 7.     The plot for $q_B=0.2$ and $L=10$ of the states which for
$q\rightarrow 1$ behave as $4A(q).$ The states in $l=1$ are marked by
triangles, in $l=2$ by crosses and in $l=3$ by pentagons. The smooth
curves are the theoretical curves~\masdisp~for particles 3 and 4 for $l=1$
and $L\rightarrow \infty.$ Note that these two curves cross near $P=1.8\pi$.

\vfill                 
\eject

\listrefs

\vfill\eject

\bye
\end